\documentstyle[graphicx,preprint,prb,aps]{revtex}
\tightenlines

\begin{document}
\draft
\title{Finite size effects in adsorption of helium mixtures by alkali
substrates}
\author{M. Barranco,$^1$ M. Guilleumas,$^1$ E.S. Hern\'andez,$^2$
R. Mayol,$^1$ M. Pi,$^1$ and L. Szybisz$^{2,3}$}
\address{$^1$Departament ECM,
Facultat de F\'{\i}sica, Universitat de Barcelona, E-08028 Barcelona,
Spain
}
\address{$^2$Departamento de F\'{\i}sica, Facultad de Ciencias Exactas y Naturales,
Universidad de Buenos Aires, 1428 Buenos Aires, and Consejo Nacional de
Investigaciones Cient\'{\i}ficas y T\'ecnicas, Argentina}
\address{$^3$ Departamento de F\'{\i}sica, CAC,
Comisi\'on Nacional de Energ\'{\i}a At\'omica, 1429 Buenos Aires, Argentina}
\date{\today}

\maketitle

\begin{abstract}

We investigate the  behavior of   mixed
$^3$He-$^4$He droplets on alkali surfaces at zero temperature, within the frame
of Finite Range Density Functional theory.  The properties of
one single $^3$He atom on $^4$He$_{N_4}$
droplets on different alkali surfaces are addressed, and the energetics and
structure of $^4$He$_{N_4}+^3$He$_{N_3}$ systems on Cs surfaces, for
nanoscopic  $^4$He drops, are analyzed through the solutions of the mean
field equations
for  varying number $N_3$ of $^3$He atoms. We discuss the size effects
on the single particle spectrum of $^3$He atoms and on the  shapes
of both helium distributions.

\end{abstract}



\section{Introduction}

The physics of wetting by quantum fluids, particularly $^4$He, $^3$He
and isotopic mixtures,  received considerable attention from both
experimentalists and theoreticians along the last decade, with
experiments mostly pointing  at  measuring adsorption isotherms,
determining  the interfacial surface tensions, establishing the wetting
temperature and constructing phase diagrams for the mixed systems.
A geometrical,   visible  parameter  that  characterizes  the  wetting
behavior of  a liquid-substrate combination at temperature  T is the
contact angle of a macroscopic  drop or wedge on a planar surface.
The contact angle of liquid  $^4$He on Cs has been measured by
at least three different groups,\cite{klier95,rolley97,ross98,klier98}
and although the reported values do not coincide -a result which among
other effects, could be attributed  to the preparation of the surface-
they  are fully  consistent with the helium inability to wet this weak
adsorber. \cite{les03} By contrast,   $^3$He has
been theoretically shown  to be a universal  wetting
agent\cite{pricaupenko94a}, a  prediction which was verified shortly
after, as
isotherms  of $^3$He  on Cs  were measured that display neat
prewetting jumps down to temperatures around 0.2 K.\cite{ross97}

  It  is well-known  that  $^3$He atoms
admixed into bulk liquid $^4$He  or $^4$He films at low T's
populate a twodimensional (2D) homogeneous
layer of Andreev states on  the free surface\cite{edwards78}
(see also  Ref. \onlinecite{hallock00}   and Refs. therein for a
recent review). In this case, the Andreev state originates in the
presence of a broad surface at the $^4$He
liquid-vapor interface and  in the mutual interaction  between the
isotopes, and corresponds to the ground state (gs) of a
discrete spectrum of states
which become progressively localized towards the interior of the film.
This structure  appears as well in density profiles of
mixed   clusters.\cite{barranco97,pi99}  It  was  earlier
proposed\cite{pav91,pavloff91b}
that   if   $^4$He lies  on a weak adsorber,  Andreev-like states could be
expected at the liquid-substrate   interface.
Theoretical anticipations  of the  wetting behavior of  mixtures rely
heavily  on  the  expected  reduction of  the  liquid-vapor  surface
tension of  the mixture with respect  to that of pure  $^4$He, due to
the  Fermi pressure  of  the Andreev  surface  layer.\cite{pettersen93}
This  effect, actually  observed in  Ref.\onlinecite{ross95},
reduces the contact angle
at a  given temperature, thus permitting  lower wetting temperatures
as the $^3$He concentration  increases.
On these grounds, Pettersen
and  Saam\cite{pettersen93}   predicted  reentrant  wetting of  helium
mixtures  on  Cs,  a phenomenon  measured  shortly after  by  Ketola
{\it  et  al}.\cite{ketola93}
The complete  phase diagram for wetting
of helium  mixtures on alkali metal substrates  derived in
Ref. \onlinecite{pettersen95}
offers a wide gallery of phenomena
which  includes   prewetting,  isotopic  separation,   triple  point
dewetting  and $\lambda$-transitions  in the  solution, in  the full
range   of $^3$He concentrations.  Subsequent   experimental  work
discovered a  rich pattern of  wetting by mixtures,  which includes,
i.e.,  dewetting  transitions  near coexistence.\cite{ross96}   More
recently,  detailed measurements  of  the contact  angle for  dilute
helium mixtures have been reported;\cite{klier99} an  analysis of the data
showed that  the large  values of these  angles are   consistent
with the presence of single-particle (sp) states of the $^3$He atoms,
	together with ripplons,
at the liquid-solid interface.

The first calculation of   density  profiles of  finite
droplets of $^4$He  on  Cs, at zero temperature, 
was forwarded by Ancilotto {\it et al}.\, within the framework
of Finite Range Density Functional Theory (FRDF) 
\cite{ancilotto98}, using the FRDF of Ref. \onlinecite{dalfovo95}.
More recently, we have presented  calculations of
nanoscale mixed $^3$He-$^4$He droplets  on  Cs,\cite{mayol03}
(hereafter, referred to as I) using a previously
derived FRDF for mixtures.\cite{barranco97,mayol01} An important
prediction in
this paper is the existence of  edge states,  the lowest-lying
$^3$He sp  bound states, which are essentially onedimensional (1D)  and
localize around the contact line of the $^4$He drop
-in contrast to the 2D surface Andreev states.

The purpose of this work is to examine in more detail
the  behavior of  mixed helium
droplets on alkali surfaces, at T=0. For this sake, in Sec.
II we shortly review the current FRDF formulation,
and in Sec. III we analyze the
energetics and structure of a single $^3$He atom on $^4$He$_{N_4}$ droplets
on alkali planar surfaces, for varying number $N_4$
of $^4$He atoms.
In Sec. IV we discuss the solutions of the mean field equations
for mixed $^4$He$_{N_4}$+$^3$He$_{N_3}$ drops on Cs
with varying number $N_3$ of $^3$He atoms.
 Emphasis here lies
on the features of the sp spectrum of $^3$He atoms  predicted by the
solution of the Kohn-Sham (KS) equations derived from the FRDF, as well as
on the shapes of both helium distributions. Our conclusions and
perspectives are summarized in Sec. V.

\section{Density functional description of pure and mixed helium clusters on
adsorbing substrates}

The FRDF for helium mixtures adopted in this work is the same as in
Ref. \onlinecite{mayol01}, of the general form
\begin{equation}
E = \int d {\bf r}\, \left[{\cal U}_3(\rho_3, \tau_3, \rho_4)
+{\cal U}_4(\rho_4, \tau_4)
+{\cal U}_{34}(\rho_3, \rho_4)\right] \;\;\; ,
\label{eq2}
\end{equation}
where $\rho_i({\bf r})$,  $\tau_i({\bf r})$ are respectively  the particle
and the kinetic energy densities for $i$ = 3, 4.
The detailed form of  the FRDF and values of the force coefficients
have been given in Ref.\onlinecite{barranco97}, with the only
changes reported in Ref. \onlinecite{mayol01} which consist,
on the one hand, in the neglect of
the nonlocal gradient correction to the kinetic energy
term  in Ref. \onlinecite{dalfovo95},
and on the other hand, in the choice of the
suppressed Lennard-Jones core as
proposed earlier.\cite{dupontroc90} More
specifically, for $^3$He-$^3$He, $^3$He-$^4$He and $^4$He-$^4$He
interactions, the Lennard-Jones screened interaction is written as

\begin{equation}\label{eq3}
V_{LJ}(r) =
\left\{
\begin{array}{lr}
4 \epsilon_{ii} \left[ \left( \sigma_{ii}/r \right)^{12} -
\left(\sigma_{ii}/r\right)^6 \right] &
\quad\rm{if} \ \ r \geq h_i \\
V_0 (r/h_i)^4  &
\quad\rm{if} \ \ r \leq h_i \ , \end{array}
\right.
\end{equation}
with $\epsilon_{44}=\epsilon_{33}=\epsilon_{34}=10.22 \,{\rm K}$,
$\sigma_{44}=\sigma_{33}=2.556 \,{\rm \AA}$,
$\sigma_{34}=2.580 \,{\rm \AA}$,  hard-core radii $h_4$= 2.359665 $\rm
\AA$, $h_3$ = 2.356415
 $\rm \AA$ and $h_{34}$ = 2.374775 $\rm \AA$, and with
$V_0$  the value of the corresponding 6-12 potential at $r=h_i$.
These values have
been fixed so that the volume integrals of the interactions
$V_{LJ}$ coincide with the original ones in
Refs. \onlinecite{barranco97,dalfovo95}.

In the current geometry,
for axially symmetric droplets, we have:

\begin{eqnarray}
\rho_3(r, z) &=& 2\;\sum_{n\, l} 
\left|\Psi_{nl}(r, z, \varphi)\right|^2
\\
\tau_3(r, z)&=& 2\;\sum_{n\, l} \left|\nabla \Psi_{nl}(r, z, \varphi)\right|^2
\\
\tau_4(r, z) &=& \frac{1}{4}\,\frac{|\nabla \rho_4(r, z)|^2}{\rho_4(r, z)}
\;\;\; ,
\end{eqnarray}
where

\begin{equation}
\Psi_{nl}(r, z, \varphi) = \psi_{nl}(r, z) \,
\frac{e^{\displaystyle i l \varphi}}{\sqrt {2 \pi}}
\end{equation}
denotes the sp wave functions (wf's) of the $^3$He atoms, and
$l$ denotes the projection of the sp orbital angular momentum
on the $z$ axis. Since we shall only address  spin saturated systems, the sp
energy levels
are either twofold ($l=0$) or fourfold ($l\neq0$) degenerate.

The Euler-Lagrange (EL) equations arising from functional differentiation
of the density functional in Eq. (\ref{eq2}) give rise to an
integrodifferential
coupled set
\begin{eqnarray}
\left[-\frac{\hbar^2}{2 m_4}\,\nabla^2  +
V_4\left(\rho_3, \tau_3\, \rho_4\right)\right]\,\sqrt{\rho_4}
=\mu_4\,\sqrt{\rho_4}
\label{eq4}
\\
-\nabla \left(\frac{\hbar^2}{2 m^*_3}\,\nabla \Psi_{nl}\right)  +
V_3\left(\rho_3, \tau_3, \rho_4\right)\, \Psi_{nl}
=\varepsilon_{nl}\,\Psi_{nl}
\label{eq5}
\end{eqnarray}
In Eq. (\ref{eq4}), $\mu_4$ is the chemical
potential that enforces particle number
conservation in the $^4$He drop,
and expression (\ref{eq5}) represents the Kohn-Sham (KS) equations
for $^3$He atoms in the presence of their mutual interaction, their coupling
to the $^4$He cluster and the potential $V_s(z)$ generated by the
alkali substrate filling the the $z \leq 0$ half-space. For
$z \geq 0$, the mean fields $V_i(r, z)$ include
$V_s(z)$  chosen as the Chizmeshya-Cole-Zaremba (CCZ) potentials.
\cite{chizmeshya98}
%
The energies ${\cal E}_i$, $i$= 3, 4, of one helium atom in the substrate,
 obtained with the CCZ potentials, are shown in Table \ref{Tab1}.

Eqs. (\ref{eq4}) and (\ref{eq5}) have been discretized using
7-point formulae and solved
on a 2D $(r,z)$ mesh.
We have used sufficiently large boxes with spatial
steps $\Delta r = \Delta z = h_4/ 12 \sim 0.197 {\rm \AA}$. As indicated
in some detail in Ref. \onlinecite{bar03}, we have employed an
imaginary time method to find the solution of these equations written
as imaginary time ($\tau$) diffusion equations. After every $\tau$-step,
the $^4$He density is normalized to $N_4$, whereas the $^3$He wf's are
orthonormalized using a Gram-Schmidt scheme. To start the iteration
procedure, we have used the halved density of a $^4$He$_{2N_4}$ cluster
obtained form a spherically symmetric FRDF code, and random numbers to
build the $^3$He wf's $\psi(r,z)$. This leaves little room to introduce any
bias in the final results.

\section{One $^3$He atom on $^4$He clusters spreaded on alkai surfaces}

The first problem to consider is the sp spectrum
for one single $^3$He atom in the field of a $^4$He$_{N_4}$ cluster
on an alkali surface.
As an example, we
solve the corresponding Schr\"odinger equation for the $^3$He atom
for $^4$He drops with $N_4$ = 20 and 100 on a Cs
surface. In Fig. \ref{fig1} we show
the sp level scheme $\varepsilon_{nl}$ in terms of squared angular momentum
$l$, up to  ${\cal E}_3$ = -3.13 K
(cf. Tab. \ref{Tab1}). For comparison,
 the lowest panel displays the  sp spectrum obtained in I for
a cluster with $N_4$ = 1000.
Localized states with energy higher than -3.13 K are meaningless,
as they would
be an artifact of the calculation, carried out in a large but finite box.
Indeed, on any substrate, $^3$He atoms with energies higher than the
corresponding ${\cal E}_3$, would
prefer to leave the $^4$He neighborhood and occupy the lowest lying
sp state of the alkali surface potential.
One can see from Fig. \ref{fig1} that whereas there are many $^3$He
sp states below ${\cal E}_3$ for $N_4=1000$ (notice that  only the
lowest lying ones are shown),
their number decreases with decreasing $N_4$. In particular, for $N_4=20$
only the states with $n=1$ and $l=0$ to 3 ($s, p, d$, and $f$ states in
spectroscopic notation), and the 2$s$ state are bound to the $^4$He droplet.

In I we have shown that, for large $N_4$ values,
 states $(nl)$ with $n=1$ and $l=0, 1, 2, \ldots$
are distributed into a rotational band. This means, on the one hand, that
their sp energies lie on a perfect straight line as  functions of $l^2$
as seen in the bottom panel of Fig. \ref{fig1}, and on the other hand,
that their
probability distributions $|\psi_{1l}(r,z)|^2$ are sensibly identical to
$|\psi_{10}(r,z)|^2$. Depending on $N_4$, the same may happen, to some
extent, to  $(2l)$ states
with  $l=0, 1, 2, \ldots$, (cf. Fig. \ref{fig1}).
For $N_4$ = 20, the small size of the $^4$He host cluster
does not allow the energy levels to group into rotational bands, even
for gs-based states $\psi_{1l}$. This is due to the fact that the wf's
$\psi_{n0}$ lie close enough to the $z$ axis to experience  centrifugal
distortion at any nonvanishing $l$. The rotational character of the gs band
$\varepsilon_{1l}$   is recovered for $N_4$ = 100, as
depicted in the middle panel of Fig. \ref{fig1}.

Contour plots of the probability densities
$\left[\psi_{n0}(x, z)\right]^2$ for $n$ = 1 to 3,
on the $(x, z)$ plane,
together with those
of the density  $\rho_4(x, z)$, are shown in Fig. \ref{fig2} for
$N_4$ = 100. It is clear that the gs $\psi_{10}$ is localized
on the circular contact line, revealing once again
the edge state reported in I, where similar plots
were presented  for  $N_4$ = 1000. In that work, we also showed that
for $n>1$, the probability densities display several
fringes on the surface of the $^4$He cluster; a similar
pattern occurs for $N_4$ = 100.
This is not the case if $N_4$ is as low as 20, since although the
edge state is still present,   the small host
immediately pulls out the $^3$He probability density into unbound
configurations (cf. upper panel in Fig. \ref{fig1}).
We show in Fig. \ref{fig3}
contour plots of the probability densities
$\left[\psi_{1l}(x, z)\right]^2$ for $l=0$, 1, and 2
(hereafter all the contour plots figures are drawn on the $y=0$ plane),
together with those
of the density  $\rho_4(x, z)$, for  $N_4$ = 20.

Effective masses $m^*_{n0}$, defined\cite{pricaupenko94a}  as
the state averages of the parametrized local prefactor of the $^3$He kinetic
energy in the density functional Eq. (\ref{eq2})

\begin{equation}
\frac{1}{m^*_{nl}} = \int d{\bf r} \;
\frac{\left[\psi_{nl}(r, z)\right]^2}
{m^*_3[\rho_4(r,z), \rho_3(r,z)]}
\label{eq7}
\end{equation}
are displayed in Table \ref{Tab2}.
The rotational character of the gs band $\varepsilon_{1l}$
if $N_4=$ 100, allows one to fit these sp energies
to a law

\begin{equation}
\varepsilon_{1l}= \varepsilon_{10} + \frac{\hbar^2 l^2}
{2 m^*_{10} R^2_{10}}
\label{eq8}
\end{equation}
with $\varepsilon_{10}$ =  -4.52 K and regression unity. From
$m^*_{10}$ = 1.18 $m_3$, we obtain a gs radius  $R_{10}$ = 15.5 $\rm \AA$,
which sensibly coincides with the geometrical radius of the droplet
at a $^4$He density of about 10$^{-2} {\rm \AA}^{-3}$,
as viewed in the  bottom panel of Fig. \ref{fig2}. 

The appearance of the $^3$He level structure built on
these spreaded droplets is
quite distinct. Within an strictly independent particle model,
one may see from Fig. \ref{fig1} that a fairly large
amount of $^3$He is needed before 2$l$ states start being occupied.
For $N_4$=100, only after filling the $1h\, (l=5)$ state, i.e., $N_3=22$, the
state 2$s$ becomes occupied, and for $N_4$=1000,
only after filling the $1k\, (l=8)$ state, i.e., $N_3=34$.
For this reason, the KS results discussed in Sec. IV refer to
rather small $^4$He droplets. Otherwise, the number of $^3$He sp
orbitals to  compute, in order to see a sizeable effect due to
the presence of this isotope,  becomes prohibitive.

It is also interesting to see how the number of $\psi_{n0}$ localized
bound states,
with energies below the ${\cal E}_3$ value in Table \ref{Tab1}, varies
with $N_4$
and the alkali substrate. In particular, for Na and Li we have found
only one single
$l=0$ bound state, namely the edge state $\psi_{10}$, for $N_4$ values from 20 to
3000. This is at variance with the results found for K, for which,
similarly to Cs, the droplet with
$N_4$=20 can only sustain one $s$ state,  whereas the drop with
$N_4$=100 and larger may sustain  $n=1$, 2 and 3 $s$ states.
We show in Fig \ref{fig4} the contour plots of the probability densities
$\left[\psi_{n0}(x, z)\right]^2$ for $n$ = 1 to 3,
together with those
of the density  $\rho_4(x, z)$,  for
$N_4$ = 3000 on K. As in the case of Cs\cite{mayol03}, the 
probability densities of the excited states 
display peaks on the upper surface of the drop.

Contour plots of the probability density $\left[\psi_{10}(x, z)\right]^2$,
together with those
of the density  $\rho_4(x, z)$,  for
$N_4$ = 3000 on Li and Na, are shown in Fig. \ref{fig5}.
The fact that $^4$He drops on Na and Li only support one bound $s$
state, namely that surrounding the contact line,
together with the rather weak $l$ dependence of the sp energies
in the $\psi_{1l}$ rotational band for large $N_4$,  makes
the $^3$He$_{N_3}$ rings  hosted by these droplets, physical
realizations of a neutral Luttinger liquid.\cite{lut62,hal81}

Figure \ref{fig6} shows the energy $\varepsilon_{1s}$ of the edge state,
as a function of  $N_4$ for Cs, K, Na, and Li substrates, with
$\varepsilon_{1s}(N_4=0) \equiv {\cal E}_3$. It should be kept in mind
that, as shown  in Ref. \onlinecite{bar03},
 in the current FRDF description,
Li, Na and K are wetted by $^4$He, whereas Rb (not discussed here) and Cs
are not. To establish a connection with the situation in films, for a given
$N_4$ droplet  we define

\begin{equation}
n_4(r) =2\,\pi\,\int dz\,\rho_4(r, z)
\label{eq100}
\end{equation}
and in view of the analysis carried out in
Ref. \onlinecite{bar03}, it is safe to  consider that $n_4(0)$ for
the largest
droplets $N_4=3000$ represents the prewetting coverage $n_c$ of a
$^4$He film on the given
adsorber. 
This yields $n_c=0.467 \,{\rm \AA}^{-2}$ for K,
$n_c=0.140 \,{\rm \AA}^{-2}$ for Na, and
$n_c=0.056 \,{\rm \AA}^{-2}$ for Li.
We have obtained the structure of films in the vicinity of $n_c$,  working
out the first three  $^3$He sp states as illustrated in
Figs. \ref{fig7} and \ref{fig8}. 

In Fig. \ref{fig7} we plot the $^4$He density as a function of
the perpendicular distance $z$,  for films on K, Na and
Li, computed at the corresponding $n_c$, together
with the mean field for one $^3$He atom, the gs wave function
$\psi_{0}(z)$ and the first excited state wf $\psi_{1}(z)$.
In Fig. \ref{fig8} we
display the energies $\varepsilon_{n}$, $n$ = 0 to 2 
(see also Fig. \ref{fig7}) for $^4$He
coverages around $n_c$ on the same three substrates, with the gs
energy ${\cal E}_3$ of the $^3$He atom on the substrate shown
as a dotted line. The heavy dot 
indicates their corresponding $n_c$.\cite{nota}
As seen in
Fig. \ref{fig7}, the wave functions follow the
trend anticipated in Ref. \onlinecite{pavloff91}:  the gs of a
$^3$He impurity in a $^4$He film  remains mostly concentrated inside the Li
monolayer, and is localized
at the liquid-vapor interface if the film thickness is above one layer
(K and Na). 

The energetics of the impurity shows an interesting evolution
across the prewetting transition. In the case of K, 
the weakest adsorber of this series, for the range of $n_4$ 
values here considered,
the film spectrum is essentially constant with coverage, and
for all three adsorbers
the first {\em excited} state appears above the gs of the $^3$He
atom on the substrate.  In contradistinction,
at the prewetting density $n_c$
the energy of the Andreev state is {\em below} ${\cal E}_3$ for K,
and {\em above} it for Na and Li.
 We appreciate then
that an important finite size effect for $^4$He clusters on these later two
adsorbers is  the generation of  the edge state for the $^3$He impurity,
which is favored with respect to binding to the substrate, thus
permitting  the existence of {\em mixed} drops 
of both helium isotopes on these adsorbers.
It may also be inferred from Fig. \ref{fig8}
the different behaviour of adsorbers like K and weaker, as compared to
Na adsorbers and stronger. Whereas in the former case, after
filling the edge-like states, $^3$He atoms start populating
Andreev-like states covering the cup of the $^4$He droplet, in
the latter, once the edge-like states are filled, $^3$He 
spreads on the alkali surface.

\section{Mixed clusters on Cs}

We  now  solve the  coupled EL/KS
equations (\ref{eq4}) and (\ref{eq5}) for  $N_4$ = 20 and 100, and for
several $N_3$ values, on a Cs substrate. Fig. \ref{fig9}
displays the variation
of the lowest KS sp energies $\varepsilon_{nl}$ as functions of $N_3$ for the
case $N_4$ = 20, with heavy dots indicating the  Fermi energy/chemical potential
$\mu_3$ of the given
configuration. 
As seen in this figure, the
chemical potential $\mu_3$ is a -nonmonotonically- increasing function of
the system size. This  trend coincides with that encountered in
$^3$He, either pure\cite{pricaupenko94a} or dissolved in  thick
$^4$He films.\cite{pav91,pavloff91b}
Figure \ref{fig9} also shows that for $N_4$=20, values of
$N_3=6$, 10, 16, and 30 yield strong shell closures, and they
are thus the magic numbers for the fermion component of
a $^4$He$_{20}$+$^3$He$_{N_3}$ droplet on Cs -these numbers depend
on the alkali substrate and on $N_4$-. Weaker shell closures
also appear for $N_3=2$, 20, 24, 34, and 38.

Contour  plots of the particle densities 
$\rho_3(x, z)$ and $\rho_4(x, z)$ are shown in Fig.
\ref{fig10} for $N_4=20$,  and in Fig. \ref{fig11} for $N_4=100$,
and two values of $N_3$.
As we analyze these patterns, several observations may be put forward. On the
one hand, we verify the spreading tendency of the density profiles
$\rho_3$  with growing $N_3$, as well as a tendency to
cover the horizontal base and  show some slight dilution inside
the bulk of the host.
It is also apparent from these figures that
 $^3$He atoms in Andreev surface states do coat the $^4$He cap.
 On the other hand, the shape of the $^4$He droplet is rather
insensitive
to  the presence of the intruder atoms, whose outwards spreading does
not induce
splashing of the $^4$He cluster on the substrate surface as well.

The behavior of the $^3$He density should be interpreted
together with examination of the KS level scheme in Fig. \ref{fig9}:
an energetically based criterion
for spreading of mixtures of $(N_3, N_4)$ atoms on any substrate is given by the
constraint  $\mu_3(N_{3c}, N_4) = {\cal E}_3$,
which defines a critical  size $N_{3c}(N_4)$ above which
it is energetically more convenient, for an extra $^3$He atom, to adhere to the
flat substrate than  to the $^4$He drop. As seen in Fig. \ref{fig9},
for $N_4$=20, $N_{3c}$ is about 44.

Finally, a word of caution should be spoken with respect to the contact angle,
for whose determination in large, macroscopic droplets, a procedure
has
been developed in Ref. \onlinecite{ancilotto98}. As discussed in I, due to the
fact that nanoscopic $^4$He droplet
densities present large stratification near the
substrate,  the contact angle cannot be established by
`visual inspection' of the equidensity lines. Yet, even at a
qualitative
level of description, it is clear from the results here presented
that addition  of $^3$He atoms affects the contact
angle. To illustrate this assertion further, in
Fig. \ref{fig12} we show the
contour plot of the total $^4$He+$^3$He density
for $^4$He equidensity lines about 0.011 ${\rm \AA}^{-3}$,
roughly half  the bulk  $^4$He density,
for  $N_4=20$ and 100 and several $N_3$ values.
Although the small size of these droplets prevents us from  drawing
quantitative conclusions, this figure indicates that the height of the cap
changes less than the radius of its base, thus suggesting overall
flattening of the combined density and a decrease of  the contact
angle, as expected according to both theory and experiment carried
upon macroscopic samples.

\section{Summary}

In our previous work here denoted as I, we have shown that in addition to the
well known Andreev states, nanoscopic $^4$He drops on Cs substrates
host a new class of $^3$He sp states, edge states that generate a
1D ring of $^3$He atoms along the contact line.
In this work we have carried a deeper investigation of size effects in
mixed droplets on planar alkali surfacess, that enables us to assert that
these edge states are a
definite consequence of the particular combination of the confinement
provoked by the adsorbing field,
selfsaturation of the helium system  and mutual interaction between
isotopes. In fact, these
localized states appear in all alkali substrates and for all $^4$He
clusters, although their 1D nature becomes more evident either above
nanoscopic host sizes or for the strongest confining potentials. We
are able to compare the energy spectrum of single $^3$He impurities
 in $^4$He clusters with similar spectra on films, and found that
while the evolution of the lowest lying sp states with film coverage
near the prewetting transition may prevent solvation of a $^3$He atom
in a film on a strong adsorber such as Na or Li, its gs energy being
unfavorable compared with binding
 to the substrate, the breaking of translational symmetry  introduced
by the finite size of clusters does change this trend, giving rise to
an energetically favored edge state which remains localized around the
drop.  We also show   that increasing $N_3/N_4$ ratios
enlarge the tendency of the mixed cluster to spread outwards on the
surface; explicit KS calculations of the energy spectrum of added
$^3$He atoms and examination of the Fermi energy as a function of
$N_3$ support the quantummechanical interpretation for the onset
of wetting by mixtures of helium fluids, as driven by the energetic
benefit of  abandoning the $^4$He and adhering instead to the substrate.

\acknowledgements

This work has been performed under grants BFM2002-01868  from
DGI, Spain, 2001SGR00064 from Generalitat of Catalonia,
EX-103 from University of Buenos Aires, and
PICT2000-03-08450 from ANPCYT, Argentina. E.S.H. has been also funded
by M.E.C.D. (Spain) on sabbatical leave.

\begin{table}
\caption{
Energies ${\cal E}_3$ and
${\cal E}_4$
(K) of one $^3$He and one $^4$He atom on different alkali substrates.
}

\begin{center}
\begin{tabular}{|c||c|c|c|c|c|}
  &  Li   &  Na  &  K   &  Rb  &  Cs  \\
\hline
${\cal E}_3$ &  -9.81  &  -6.41 &  -3.74 &   -3.31 &  -3.13 \\
${\cal E}_4$ &   -10.70  &  -7.08 & -4.20 &  -3.72 &  -3.53  \\
\end{tabular}
\end{center}
\label{Tab1}
\end{table}

\begin{table}
\caption{
Effective masses $m^*_{n0}$ (in units of $m_3$) for single $^3$He atoms
in He$_{N_4}$ clusters on Cs.
}

\begin{center}
\begin{tabular}{|c||c|c|c|c|c|c|}
$N_4$       &  20  &  100  & 500  & 1000  & 2000  &  3000  \\
\hline
$m^*_{10} $ & 1.13  & 1.18  & 1.20  & 1.21  &  1.22  & 1.22  \\
$m^*_{20} $ & 1.10  & 1.26  & 1.29  & 1.30  &  1.31  & 1.31  \\
$m^*_{30} $ & 1.05  & 1.24  & 1.30  & 1.30  &  1.30  & 1.30  \\
\end{tabular}
\end{center}
\label{Tab2}
\end{table}

\begin{figure}
\caption[]{
Single-particle energies $\varepsilon_{nl}$ of one
$^3$He atom in a $^4$He$_{N_4}$ droplet on Cs, as functions of squared angular
momentum projection, for $N_4$ = 20, 100 and 1000.
The dotted lines have been drawn to guide the eye.
In each panel, from bottom to top, the states have principal quantum numbers
$n=1, 2, 3 \ldots$, and the dot-dashed line represents the gs energy
${\cal E}_3$ of one $^3$He atom on the Cs substrate.
 }
\label{fig1}
\end{figure}

\begin{figure}
\caption[]{
Contour plots of the probability densities
$\left[\psi_{n0}(x, z)\right]^2$  for
a single $^3$He atom in a $^4$He$_{100}$ droplet on Cs, together
with  the density $\rho_4(x, z)$ (bottom panel). 
In the case of $^4$He, the equidensity lines correspond to
$10^{-4}, 10^{-3}, 2.5 \times 10^{-3}, 5 \times 10^{-3},
7.5 \times 10^{-3}, 10^{-2}, 2 \times 10^{-2}$, and
$2.5 \times 10^{-2}  {\rm \AA}^{-3}$.
In the case of $^3$He, the equiprobability lines start at
$10^{-4} {\rm \AA}^{-3}$ and increase in $10^{-4} {\rm \AA}^{-3}$ steps.
}
\label{fig2}
\end{figure}

\begin{figure}
\caption[]{
Contour plots of the probability densities
$\left[\psi_{1l}(x, z)\right]^2$  for
a single $^3$He atom in a $^4$He$_{20}$ droplet on Cs, together
with the density $\rho_4(x, z)$ (bottom panel). 
The equidensity and equiprobability lines are as in Fig. \ref{fig2}.
}
\label{fig3}
\end{figure}

\begin{figure}
\caption[]{
Contour plots of the probability densities
$\left[\psi_{n0}(x, z)\right]^2$  for
a single $^3$He atom in a $^4$He$_{3000}$ droplet on K, together
with the density $\rho_4(x, z)$ (bottom panel). 
The equidensity lines are as in Fig. \ref{fig2},
and the equiprobability lines
start at
$10^{-5} {\rm \AA}^{-3}$ and increase in $10^{-5} {\rm \AA}^{-3}$ steps.
}
\label{fig4}
\end{figure}

\begin{figure}
\caption[]{
Contour plots of the probability density
$\left[\psi_{10}(x, z)\right]^2$  for
a single $^3$He atom in a $^4$He$_{3000}$ droplet on Li, together
with the density $\rho_4(x, z)$ (bottom panels), and on Na (top panels).
In the case of $^4$He, the equidensity lines are as in Fig. \ref{fig2},
and in the case of $^3$He, the equiprobability lines correspond to
$10^{-8},  10^{-7}, 10^{-6},  10^{-5},
2 \times 10^{-5}, 2.5 \times 10^{-5}$ and
$5 \times 10^{-5}  \, {\rm \AA}^{-3}$.
}
\label{fig5}
\end{figure}

\begin{figure}
\caption[]{
Energy $\varepsilon_{10}$ of the edge state (K) as a function of $N_4$.
The dotted lines have been drawn to guide the eye.
The $y$ axis is broken below $\sim -7.5$ K.
}
\label{fig6}
\end{figure}

\begin{figure}
\caption[]{
The $^4$He density at the prewetting coverage $n_c$ on Na and Li
substrates, and at $n_4=0.40$ ${\rm \AA}^{-2}$ for K,
together with the mean field $V_3$ experienced by a
$^3$He impurity, and the gs wave function $\psi_0(z)$ and first
excited wf $\psi_1(z)$ in that field.
}
\label{fig7}
\end{figure}

\begin{figure}
\caption[]{
The  sp energies $\varepsilon_{n}$, $n$ = 0, 1, and 2,
for a $^3$He impurity in $^4$He  films at coverages around $n_c$ on K,
Na and Li. The dotted line indicates  the gs energy ${\cal E}_3$ 
of one $^3$He atom on the
corresponding substrate, and the heavy dot points to  
the coverage at the prewetting jump.\cite{nota}}
\label{fig8}
\end{figure}

\begin{figure}
\caption[]{
The Kohn-Sham sp energies $\varepsilon_{nl}$
for $^3$He$_{N_3}+^4$He$_{20}$ mixed drops on Cs as functions of $N_3$.
The horizontal line indicates the  energy
${\cal E}_3$ of one  $^3$He atom in the Cs adsorbing field.
Heavy dots indicate the corresponding Fermi energy.
The lines connecting the sp energies of $(nl)$ states have been drawn 
to guide the eye.
}
\label{fig9}
\end{figure}

\begin{figure}
\caption[]{
Contour plots of the density profiles  $\rho_3(x, z)$ (top panels) and
$\rho_4(x, z)$ (bottom panels) of $^4$He$_{20}$ + $^3$He$_{N_3}$ mixed
drops on Cs, for $N_3$ = 20 (left panels) and 40 (right panels).
In the case of $^4$He, the equidensity lines are as in Fig. \ref{fig2},
and in the case of $^3$He, they correspond to
$10^{-4}, 2.5 \times 10^{-4}, 5 \times 10^{-4}, 7.5 \times 10^{-4},
10^{-3}, 2.5 \times 10^{-3}, 5 \times 10^{-3}, 7.5 \times 10^{-3}$,
and $10^{-2} {\rm \AA}^{-3}$.
}
\label{fig10}
\end{figure}

\begin{figure}
\caption[]{
Same as Fig. \ref{fig10} for $N_4$ = 100,
and $N_3=46$ (left panels), and $N_3=86$ (right panels).
The equidensity lines are as in Fig. \ref{fig10}.
}
\label{fig11}
\end{figure}

\begin{figure}
\caption[]
{Contour plots of the total $^4$He$+^3$He density
corresponding to 0.011 ${\rm \AA}^{-3}$ for the case
$N_4=20$, $N_3=20$ and 40  (top panel),
 and $N_4=100$,  $N_3=46$ and 86 (bottom panel).
The dashed lines correspond to the pure $^4$He$_{N_4}$ drop.
}
\label{fig12}
\end{figure}


\begin{references}
\bibitem{nota} 
Actually, in the case of K, we have not reached 
$n_c$ in the calculations. This is
not crucial, since for this alkali, the spectrum is insensitive to the
actual value of the coverage for values above $\sim$ 0.40 ${\rm \AA}^{-2}$,
see for example Ref. \onlinecite{pav91}.
\end{references}
\end{document}